\documentclass[aip,rsi,amsmath,amssymb,reprint]{revtex4-1}
\pdfoutput=1
\usepackage{graphicx}  
\usepackage{dcolumn}   
\usepackage{bm}        
\usepackage{amssymb}   
\usepackage{color,epsfig}
\usepackage{amsfonts}
\usepackage{booktabs}
\usepackage{amsmath}
\def\red#1{\textcolor{red}{#1}}
\def\blue#1{\textcolor{blue}{#1}}

\usepackage{soul} 

\begin{document}
\newcommand{\HH}{H$_2$}
\newcommand{\pHH}{\emph{para}-H$_2$}
\newcommand{\oHH}{\emph{ortho}-H$_2$}
\newcommand{\Eref}[1]{Eq.~(\ref{#1})}
\newcommand{\Fref}[1]{Fig.~\ref{#1}}
\newcommand{\etal}{\emph{et al.}}

\newcommand{\beginsupplement}{%
        \setcounter{table}{0}
        \renewcommand{\thetable}{S\arabic{table}}%
        \setcounter{figure}{0}
        \renewcommand{\thefigure}{S\arabic{figure}}%
     }

\def\red#1{\textcolor{red}{#1}}
\def\blue#1{\textcolor{blue}{#1}}

\widetext


\title{Unveiling shape resonances  in H  + HF collisions at cold energies.}
\author{P. G. Jambrina} \affiliation{Departamento de Qu\'{\i}mica F\'{\i}sica. University of Salamanca, Salamanca 37008, Spain.}
\email{pjambrina@usal.es}
\author{L. Gonz\'alez-S\'anchez} \affiliation{Departamento  de Qu\'{\i}mica F\'{\i}sica . University of Salamanca, Salamanca 37008,
Spain.}\email{lgonsan@usal.es}
\author{M. Lara} \affiliation{Departamento de Qu\'{\i}mica F\'{\i}sica Aplicada. Universidad Aut\'onoma de Madrid,
Madrid 28049, Spain}\email{manuel.lara@uam.es}
\author{M. Men\'endez} \affiliation{Departamento de Qu\'{\i}mica F\'{\i}sica.  Universidad Complutense.
Madrid 28040, Spain}\email{menendez@quim.ucm.es}
\author{F. J. Aoiz} \affiliation{Departamento de Qu\'{\i}mica F\'{\i}sica . Universidad Complutense.
Madrid 28040, Spain}\email{aoiz@quim.ucm.es}
\date{\today}

\begin{abstract}

Resonances are associated with the trapping of an intermolecular complex, and are characterized by
a series of quantum numbers such as the total angular momentum and the parity,
representative of a specific partial wave. Here we show how at cold temperatures the
rotational quenching of HF($j$=1,2) with H is strongly influenced by the presence of
manifolds of resonances arising from the combination of a single value of the orbital
angular momentum with different total angular momentum values. These resonances give rise
up to a two-fold increase in the thermal rate coefficient at the low temperatures
characteristic of the interstellar medium. Our results show that by selecting the
relative geometry of the reactants by alignment of the HF rotational angular momentum, it
is possible to decompose the resonance peak, disentangling the contribution of different
total angular momenta to the resonance.

\end{abstract}

\maketitle



Scattering resonances are pure quantum mechanical effects that appear whenever the
collision energy, $E_{\rm coll}$,  matches the energy of a quasi-bound state of the
intermolecular complex. \cite{L:ACP12} Unlike other quantum effects, they can be detected
straight from the experiment, for example using molecular beams
\cite{neumark_lee:jcp85,VOCAGM:S15,JBSKAGM:S20,PDRKN:A20}, where they manifest as local,
sharp maxima in either the cross section or angular distribution of the products.

From a conceptual point of view, resonances are pictured as the result of the trapping of
the intermolecular complex in potential wells after tunneling through the barrier (shape
resonance) due to the presence of quasi-bound states, or as the excitation to a state of
asymptotic higher energy, which is stabilized by the potential well (Feschbach resonance). Indeed,
a dense resonance structure is observed for complex-forming reactions, (see for example
Refs. \cite{BCJKRAB:JPCA15,JABSBH:PCCP10,LJAL:JCP15}) where the deep potential well can
stabilize a myriad of quasi-bound states. \cite{BL:JCP20}

Very recently, Perreault \textit{et al.} measured the role that the initial alignment of
HD  plays in H$_2$+ HD collisions, which in the cold energy regime, is governed by a
resonance at around 1 K. \cite{PMZ:S17,PMZ:NC18} Using quantum mechanical scattering
calculations, it was possible to assign this resonance to a single partial wave ($L$=2)
\cite{CBHG:PRL18,CB:JCP19} and, to elucidate that, for a particular combination of
initial and final states, this resonance  could be controlled, vanishing for a
suitable alignment of the HD internuclear axis.\cite{JCGBBA:PRL19}

In this manuscript, we turn our attention to the FH$_2$ system, one of the most widely
studied systems in reaction dynamics both experimentally
\cite{neumark_lee:jcp85,skodje_liu:prl2000,qiu:science2006,dong:science2010,wang:science2013,kim:science2015,YHXCWDLASZYN:NC19}
and computationally (see for example refs.
\cite{manolo:jcsft1997,alexander:jcp2000,lique:jcp2008,jalde:pccp2011,tizniti:natchem2014,SVH:PCCP19,FAC:JPCA20}).
In particular, we will focus on H+HF inelastic collisions. HF is ubiquitous in the
universe, \cite{ACWDENTD:AA2011,Indriolo_2013} and is a key tracer of molecular hydrogen
in diffuse interstellar medium. \cite{DL:MNRAS18} Here we will show that, as far as our
calculations are concerned, collisions between H and HF($j$=1,2) in the cold energy
regime are dominated by a manifold of resonances, which have a strong influence in the
thermal coefficients for the range of temperatures relevant to the study of the
chemistry in diffuse interstellar medium. We will also study the origin of these
resonances and show the extent of control that can be achieved by preparing HF with a
given internuclear axis distribution resulting from the alignment of its rotational
angular momentum. Strikingly, our calculations predict that, for a certain alignment of
the HF molecule, it is possible not only to enhance or to diminish the intensity of the resonance, but also to split the resonance peak, allowing us to
disentangle its various contributions.

\begin{figure*}
  \centering
  \includegraphics[width=1.0\linewidth]{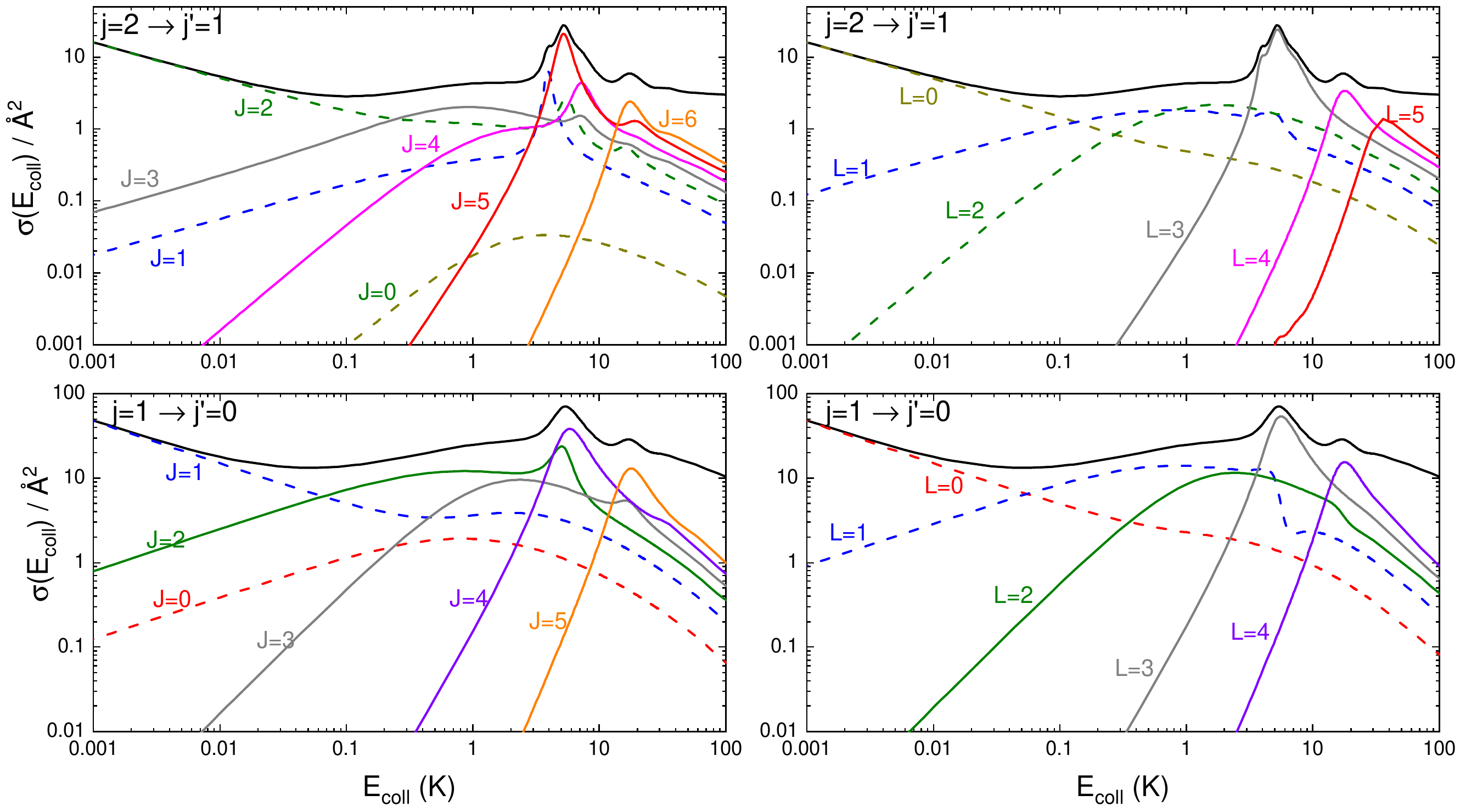}
  \caption{
Total and partial integral cross sections for the H + HF($v$=0,$j$=1,2) inelastic
collisions as a function of the collision energy. The total cross section for $j$=2
$\rightarrow$ $j'$=1 (top panels) and $j$=1 $\rightarrow$ $j'$=0 (bottom panels) are
shown as solid black lines. The contributions of each $J$ (left panels) and $L$ (right
panels) partial waves to the total cross section are also shown. } \label{Fig1}
\end{figure*}
Since the aim of this work is i) the characterization of the observed resonances, and ii)
to elucidate the extent of control that can be achieved, quantum mechanical (QM)
scattering calculations were performed in a dense grid of collision energies starting at
very low energies, $E_{\rm coll}/k_{\rm B}<$1 mK, and up to 100 K. To get an accurate
description of the dynamics, it was necessary to propagate the wave-function up to very
large distances ($6\cdot 10^4$ a$_0$), which made convenient to use the atom-rigid rotor
approximation. Calculations were performed using the ASPIN code
\cite{LOPEZDURAN2008821,GonzalezSanchez-etal:15} on the LWA-78 Potential Energy Surface,
\cite{LWLA:JCP07} which has been recently used to study H+HF collisions at higher
energies.\cite{DL:MNRAS18} To check the validity of the atom rigid-rotor approximation,
full-dimensional QM scattering calculations were carried out for a few energies using the ABC code.
\cite{SCM:CPC00} The agreement between the two sets of calculations is good (see
Fig.~S1), although the atom-rigid rotor calculations underestimates the full-dimensional
calculations at the highest energies by 20\%. Both sets of calculations show the same
overall behavior, and the main features of the full-dimensional results are well
accounted for by the rigid-rotor calculations.

 Figure~\ref{Fig1} displays the energy dependence of the rotational quenching cross
sections for HF($v$=0,$j$=1,2)+H collisions in the cold energy regime as function of the
collision energy, $\sigma(E_{\rm coll})$, in the 1\,mK--100\,K $E_{\rm coll}$ range.
Energies in this range are not sufficient to promote transitions to higher rotational
states, and, besides, the probability for the H exchange channel is
negligible.\cite{DL:MNRAS18,JGASA:JPCA19} Hence, the only possible transitions are
 $j=1\rightarrow j'=0$, and $j=2\rightarrow j'=1$, 0.
For a given $E_{\rm coll}$, $\sigma(E_{\rm coll})$ for $j=1\rightarrow j'=0$  is 3-5
times higher than for
 $j=2\rightarrow j'=1$ due to the wider gap between adjacent rotational quantum
states with increasing $j$.\cite{JGASA:JPCA19}
The Wigner regime for both channels is attained at energies below 5\,mK, where
 $\sigma(E_{\rm coll}) \propto
E_{\rm coll}^{-1/2}$. \cite{SBCEFMR:JPB00,ISCJ:NJP11}
The respective $L$ partial
cross sections, $\sigma^L(E_{\rm coll})$, and $J$-partial cross section, $\sigma^J(E_{\rm
  coll})$, are also shown in Fig.~\ref{Fig1}. As expected,
at the Wigner regime only the s-wave ($L$=0, $J$=$j$) partial wave
contributes significantly to the cross section. The corresponding Wigner limit,
$\sigma^L \propto E_{\rm
  coll}^{L-1/2}$, is also found for $L > 0$.\cite{LJLA:PRA15}

Most notably is that, at energies above the Wigner regime, $\sigma(E_{\rm coll})$ is
dominated by a narrow peak located at around 5\,K for the $j=2 \rightarrow j'=1$
transition and around 5.5\,K for the $j=1 \rightarrow j'=0$ transition. At significantly
higher energies, $E_{\rm coll} \sim$ 17\,K, a broader albeit smaller peak shows up.
Results for
 $j=2 \rightarrow j'=0$ are shown in Fig.~S2, and although $\sigma(E_{\rm coll})$ is
smaller by at least one order of magnitude, it features the same resonance peaks as  $j=2 \rightarrow j'=1$.
From inspection of  $\sigma^L(E_{\rm coll})$, it is clear that the sharp resonance peak
at 5\,mK is exclusively due to $L$=3 and that the second broader peak  can be mainly
attributed to $L$=4. Nevertheless, the analysis of $\sigma^J(E_{\rm coll})$ shows that
each peak can be decomposed in a series of maxima corresponding to different $J$s
and the same $L$.

\begin{figure}
  \centering
  \includegraphics[width=1.0\linewidth]{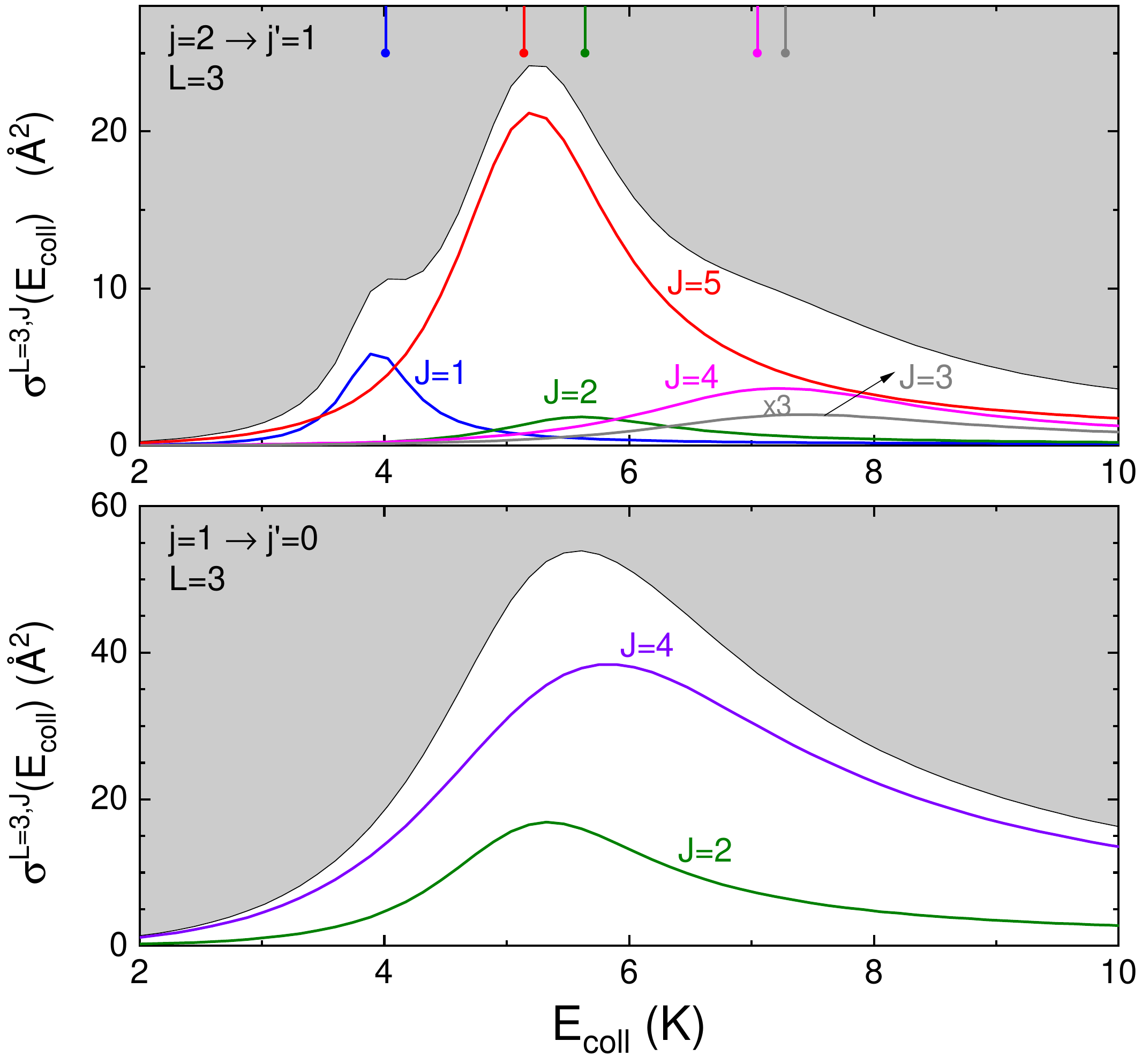}
  \caption{
$L$=3 partial cross sections for $j=2 \rightarrow j'=1$ (top panel) and $j=1 \rightarrow j'=0$
(bottom panel). The white area correspond to the total cross section
 for $L$=3, $\sigma^{L=3}(E_{\rm coll})$, while the solid lines indicate the contribution of each $J$,
 $\sigma^{L,J}(E_{\rm coll})$.  The positions of the resonance
peaks predicted by the 1D-model are shown as solid vertical lines.
 }
  \label{Fig2}
\end{figure}

A further analysis can be carried out by plotting the contributions from the different
$J$ to a given $L$, the double partial cross sections $\sigma^{L,J}(E_{\rm coll})$, as
shown in Fig.~\ref{Fig2} for $L$=3 near the resonance. For $j=2 \rightarrow j'=1$,
$J=1-5$ contribute to $L$=3 ($L - j \leq J \leq L+j$), and their respective
$\sigma^{L=3,J}(E_{\rm coll})$ show maxima that are shifted in a relatively broad
range of $E_{\rm coll}$, as shown in Table~\ref{t1}. From these peaks, $J$=1 and 5
contribute the most to the overall intensity, while the contributions of $J$=2 and 3 are
almost negligible. The shifting of the maxima for $J$=1 and 5 leads to the small shoulder
observed in the overall resonance peak. For the $j=1 \rightarrow j'=0$  transition, only
$J=$2 and mainly $J$=4 contribute to $L$=3 (due to the parity conservation). In this
case, the position of the maxima is very similar and hence there are no shoulders in the
overall $\sigma(E_{\rm coll})$.

\begin{figure}
  \centering
  \includegraphics[width=1.0\linewidth]{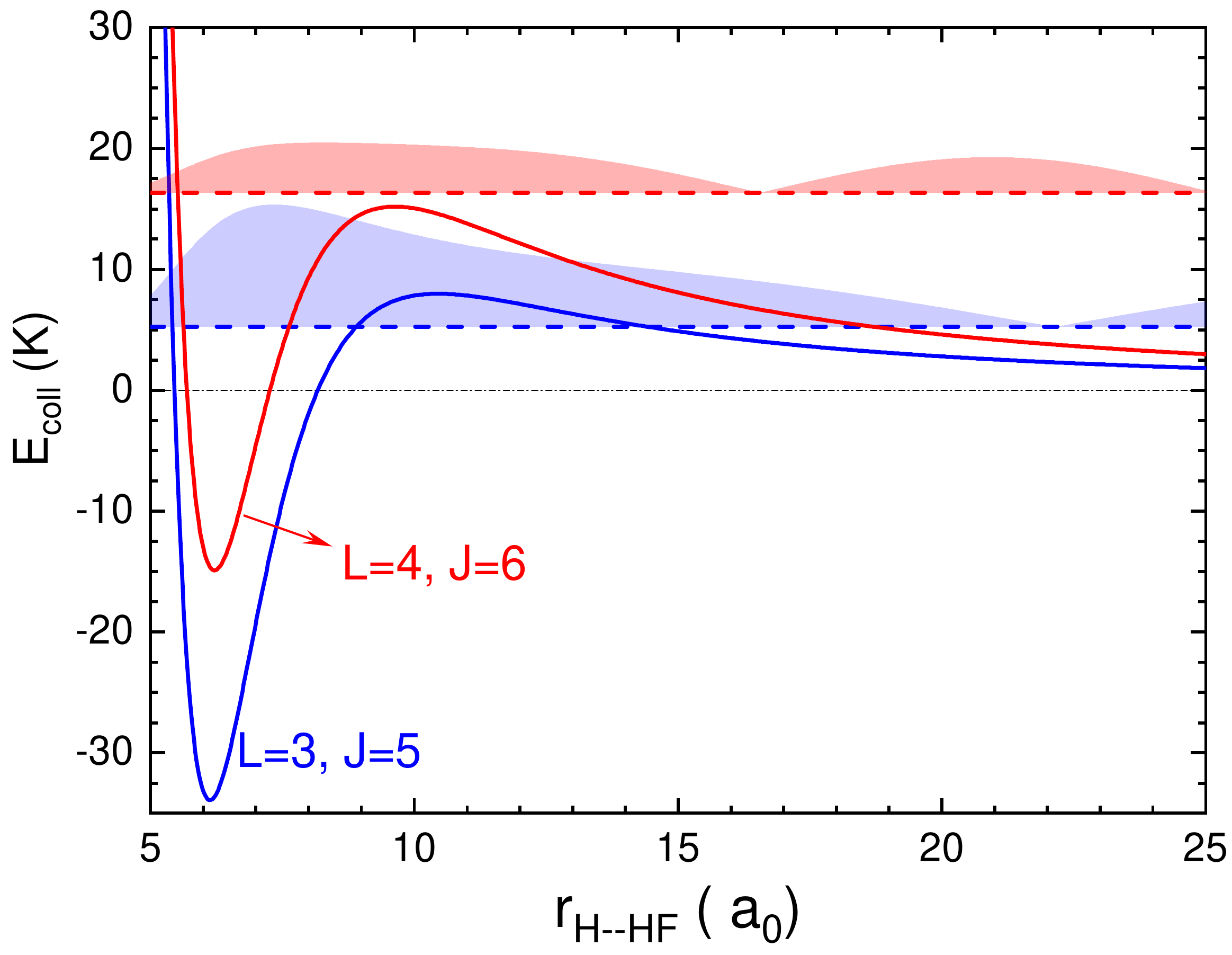}
  \caption{
One-dimensional adiabatic effective intermolecular potential as a function of the distance between H
and the center-of-mass of HF for two combinations of $J$ and $L$ of the $j=2 \rightarrow j'=1$ transition.
The energies corresponding to the maximum time delay associated with
these potentials are shown as dashed lines, along with the continuum wavefunctions. The
shaded area is proportional to the square of the wavefunction.
 }
  \label{Fig3}
\end{figure}

\begin{figure}
  \centering
\includegraphics[width=1.0\linewidth]{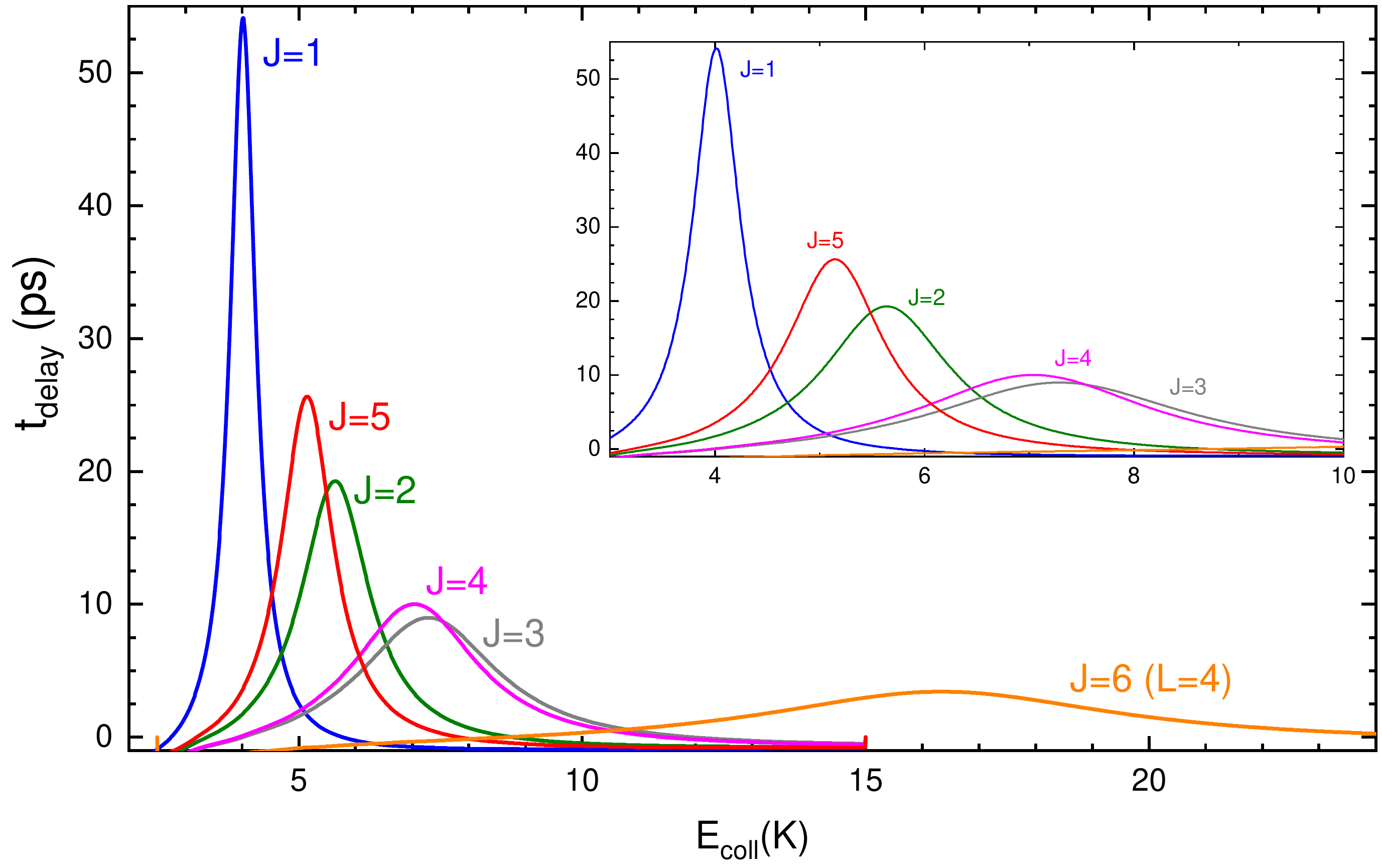}
  \caption{
$t_{delay}(E)$ as a function of the collision energy for $L$=3 manifold of resonances
for $j=2\rightarrow j'=1$ transition.  $t_{delay}(E)$ for $L$=4 and $J$=6 is also shown for the sake of comparison.
The insert shows a blow-up of the energy range where the $L=3$ resonances show up.}
\label{Fig4}
\end{figure}

To characterize the nature of the aforementioned resonances, 1D adiabatic effective
potentials for different rovibrational states have been calculated as a function of the
atom-diatom distance (see supplementary information for further details). Two of them are
shown in Fig~\ref{Fig3}. The binding character at short distances and the centrifugal
barrier are evident in the figure. By tunneling, the trapping region is accessible,
supporting quasibound states that give rise to shape resonances. The energies at which
the 1D model predicts the peaks of each ($L$,$J$) resonance are shown in Table~\ref{t1},
and in Fig.~\ref{Fig2} as vertical ticks. As can be observed, the agreement between the
energies of the resonances and the maxima of the peaks is almost perfect, allowing us to
attribute these peaks to shape resonances arising from different combinations of $J$ and
$L$. The lifetimes and line shapes associated with each resonance were also
calculated using the 1D model (Fig.~\ref{Fig4}), and the scattering probabilities.
The results are compared in Table~\ref{t1} showing that the lifetimes obtained with the
two methods are in good agreement, with the 1D model predicting slightly longer
lifetimes. The lifetimes for the $L$=3 resonances are significantly longer, revealing
that $L$=3 and $L$=4 resonances have a different character.
\begin{table}
\begin{tabular}{ |c|c|c|c|c|  }
\hline
\multicolumn{5}{|c|}{$L$=3} \\
\hline
$J$ &  $E$ (K) & $\tau$ (ps) & $E$ (K) 1D & $\tau$ (ps) 1D \\
\hline
1 & 3.9 & 11.3 & 4.0 & 14.0 \\
2 & 5.5 & 4.4 & 5.6 & 5.2 \\
3 & 7.1 & 2.4 & 7.3 & 2.6 \\
4 & 7.0 & 2.6 & 7.0 & 2.8 \\
5 & 5.1 & 4.9 & 5.1 & 6.8  \\
\hline
\multicolumn{5}{|c|}{$L$=4} \\
\hline
6 & 16.2 & 0.90 & 16.3 & 1.0  \\
\hline
\end{tabular}
\caption{Energy ($E_{\rm coll}$) and lifetimes, $\tau$, of each of the manifold of
resonances for $L$=3 and $L$=4 of the $j=2\rightarrow j'=1$ transition. The second and
third columns are, respectively, the energies of the resonance peaks and the lifetimes
obtained form the Lorentzian profiles of the collision probabilities after substraction
of the scattering background. The fourth and fifth columns are the respective data
obtained by applying the 1D model described in the Supplementary information.} \label{t1}
\end{table}
To further clarify the origin of the differences in the lifetimes of the  various
$J, L$ resonances, the effective adiabatic 1D potentials for the $L$=3--$J$=5 and
$L$=4--$J$=6 particular cases are shown in Fig.~\ref{Fig3}. Along with the potentials, we
show the energy of the quasibound states supported by these potentials (dashed lines) and
the corresponding squares of the wavefunctions. For the $L$=3 peak, the energy of the
quasi-bound state lies below the maximum of the centrifugal barrier, and it can be
properly considered as a shape resonance. However, for the $L$=4 peak, the resonance
energy lies slightly above the maximum of the barrier. As a consequence, the probability
is more evenly spread over all radial distances, and the resonance exhibits a smaller
lifetime. This kind of resonance can thus be characterized as an over-barrier (or
``above-the-barrier'') resonance, the quantum equivalent to classical-orbiting
\cite{PDRKN:A20}.

In Ref.~\citenum{JCGBBA:PRL19} it was found that the strength of the resonance peak can
be tuned by alignment of rotational angular momentum and hence by changing the
internuclear axis distribution. To check if it is also the case for H+HF collisions, we
have used the procedure outlined in
Refs.~\citenum{TutorialsbookChapter,CH:MP75,AMHKSA:JPCA05,AHJAJZ:JPCL12} to investigate
how the integral cross sections change by varying the angle $\beta$ between the
polarization vector of the radiation field, used to prepare the HF molecule in specific
rovibrational states, and the initial relative velocity vector. Each $\beta$ value
entails a distribution of internuclear axis: if $\beta$=0$^{\circ}$ collisions are
preferentially head-on, while $\beta$=90$^{\circ}$ implies a side-on geometry. The
respective cross sections will be denoted by $\sigma^{\beta}(E_{\rm coll})$. These
preparations can be contrasted with the ``isotropic'' distribution, with no external
alignment.

\begin{figure*}
  \centering
\includegraphics[width=0.80\linewidth]{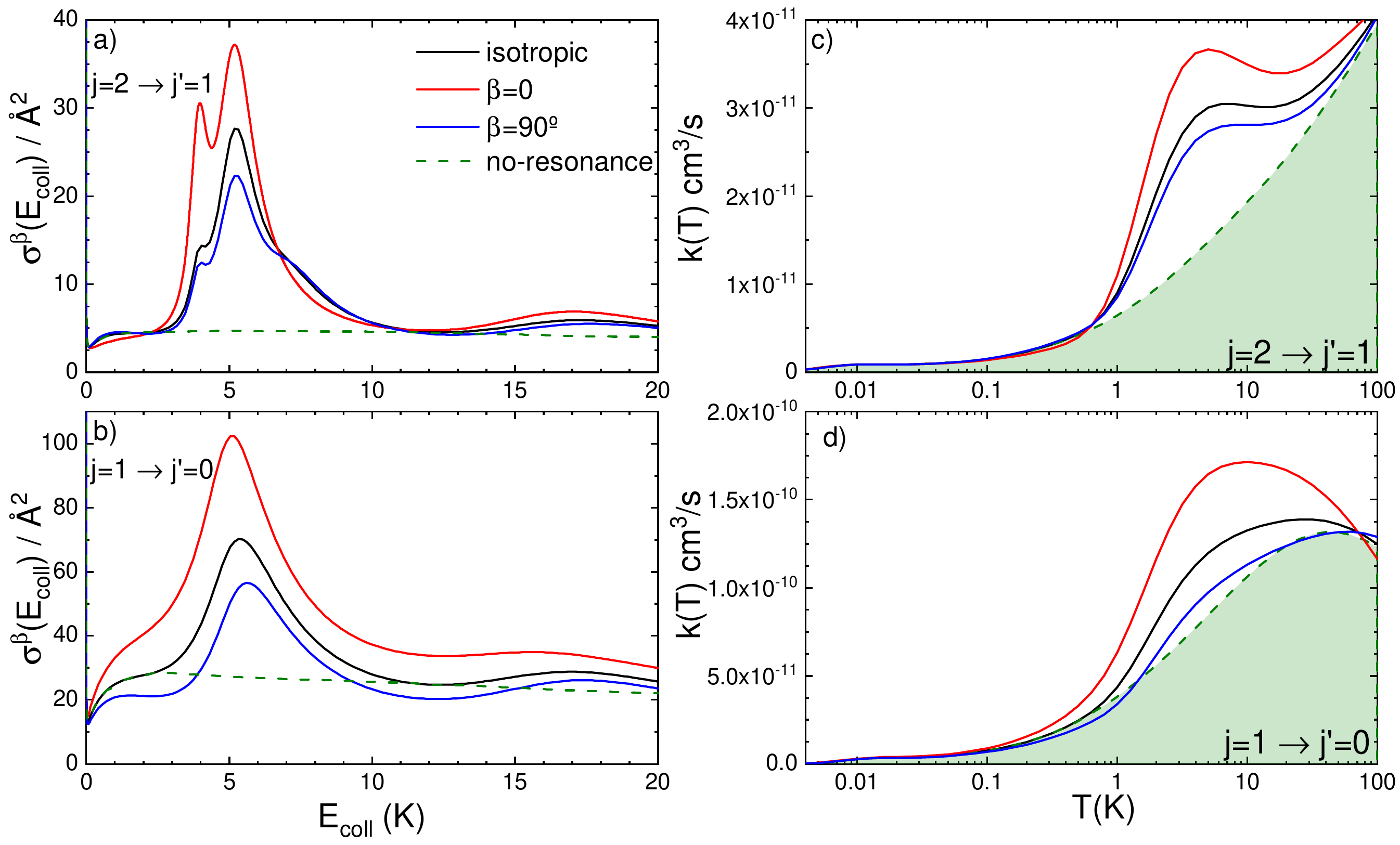}
  \caption{
Left panels: cross-sections for different preparations of HF internuclear axis,
$\beta$=0$^{\circ}$ (red line) and $\beta$=90$^{\circ}$ (blue line). The isotropic
preparation (absence of external alignment) is shown in black. The dashed lines
correspond to the background cross section when the resonance has been removed. Results
for $j=2\rightarrow j'=1$ and $j=1\rightarrow j'=0$ are shown in panels (a) and (b),
respectively. Right panels: Thermal rate coefficients (averaged over the MB distribution)
for the mentioned transitions and internuclear axis preparations. The contribution of the
background (no resonance) is shown shaded.}
\label{Fig5}
\end{figure*}

The left panels in Fig~\ref{Fig5} display the cross sections for $j=2\rightarrow j'=1$
(Fig.~\ref{Fig5}-a) and $j=1 \rightarrow j'=0$ (Fig.~\ref{Fig5}-b) in the vicinity of the
$L$=3 resonance for isotropic distribution, $\beta$=0$^{\circ}$ and $\beta$=90$^{\circ}$.
For $j=2\rightarrow j'=1$ (a) the $L$=3 resonance is significantly enhanced for head-on
($\beta$=0$^{\circ}$) encounters while its intensity decreases for side-on encounters.
Outside the resonance region, the effect of a preferential alignment is unimportant. The
most interesting feature is that the $\beta$=0 alignment is able to disentangle the peaks
for $J$=1 and $J$=5; the contribution of $J$=1, which in the isotropic case manifests as
a shoulder in the $L$=3 resonance, is enhanced to the point of splitting the original
peak in two. Since $\beta$=0 implies collisions with $\Omega$=0 exclusively, where
$\Omega$ is the projection of the total angular momentum vector onto the relative
velocity, this implies that the $J$=1 partial wave has a strong component of the
perpendicular projection. For $j=1\rightarrow j'=0$ (Fig. \ref{Fig5}-b) the cross-section
is also enhanced for $\beta$=0, and this effect is particularly prominent at the
resonance. In addition, the preference for head-on collisions is observed in a broad
range of $E_{\rm coll}$. In this case, the resonance peak does not split for any HF
preparation, as the energies corresponding to the two shape resonances contributing to
this peak are very similar. These results for both transitions evince that the
trapping of the collision complex is more efficient for head-on collisions.

Figures~\ref{Fig5} c-d show the thermal rate coefficients, $k(T)$, calculated for the two
transitions averaged over the Maxwell-Boltzmann distribution. In general, it is not
guaranteed that a resonance may influence the rate coefficients significantly. However,
in the present case, the resonance has a strong effect on $k(T)$ in the 1-50\,K
temperature range, leading to more than a two-fold increase of $k(T)$ for $j$=2
$\rightarrow$ $j'$=1 at 5 K. The different intermolecular axis preparations also have a
strong effect on the $k(T)$ in the same temperature range, with $\beta=0$ encounters
leading to the largest $k(T)$. It also worth noticing that, while for $j$=2 $\rightarrow$
$j'$=1 the $k(T)$ calculated with the resonance artificially removed rises
monotonically with $T$, for $j$=1 $\rightarrow$ $j'$=0 it starts decreasing at 40 K,
effect that is even more clear for $\beta$=0.

In summary, in this study we have demonstrated that H + HF inelastic collisions in the
1-10\,K ($E_{\rm coll}/k_{\rm B}$) energy range are dominated by shape resonances, which
are themselves formed by a cluster of resonances, each of them characterized by orbital
and total angular momentum values. We have shown that a 1-D model, based on the adiabatic
effective potentials, can predict the position of the each $L$-$J$ resonance very
accurately. Lifetimes and line-shapes of each of the resonances were determined
using the phase-shift of the 1D continuum wavefunctions. In particular, the $L$=4
resonances exhibit shorter lifetimes and can be considered orbiting (over-the-barrier)
resonances where the quasi-bound states lie slightly above the centrifugal barrier. In
spite of the relatively large time delays associated to the resonances, alignment of HF
prior the collision changes significantly both the intensity and shape of the excitation
functions, which also manifest in the thermal rate coefficients. Remarkably, for
$j=2\rightarrow j'=1$, when head-on collisions are promoted, the main resonance peak
splits in two, each of them associated to collisions with a particular value of the total
angular momentum. These results are in contrast to those found for HD + H$_2$ collisions
in the cold energy regime, for which the resonance vanishes for head-on encounters,
showing that the degree of control associated to the resonance is very sensitive of the
topology of the system. The influence of the resonance persists after the energy
averaging and it leads to up to two-fold increase of the thermal rate coefficient at
relevant temperatures of the interstellar medium where HF is ubiquitous.

%
The authors thank Prof. Enrique Verdasco for his support and help with the
calculations. Funding by the Spanish Ministry of Science and Innovation (grant
and PGC2018-096444-B-I00) is also acknowledged.
P.G.J. acknowledges funding by Fundaci\'on Salamanca City of Culture and
Knowledge (programme for attracting scientific talent to Salamanca).

\bibliographystyle{apsrev4-1}


%

\clearpage

\beginsupplement

\section{Supplementary Material}


\subsection{Scattering calculations:}

The usual time-independent formulation of the coupled channel (CC) method is used to
solve the Schr\"odinger equation, in the quantum scattering calculations of an atom with
a diatomic molecule,  as implemented in ASPIN code.\cite{LOPEZDURAN2008821}  The HF
molecule in its singlet ground state is treated as a rigid rotor, while the H atom is
considered structureless. Details of the method have been given before and discussed
recently in detail for the case of the Rb + OD$^-$/OH$^-$ system
\cite{GonzalezSanchez-etal:15} and  will not be reported here.  The parameters used for
applying the CC method are chosen for achieving numerical convergence of the final
S-matrix elements.  A maximum number of rotational channels up to $j_{max}$ = 10 has been
included in each CC calculation, where at least five channels were included as closed
channels for each collision energy, ensuring overall convergence of the inelastic cross
sections. At the lowest considered collision energies the radial integration was extended
out to $R_{max}$ =$5\cdot 10^4$ \AA.

To check the validity of the rigid rotor approximations, full-dimensional QM calculations
were also obtained using the ABC code. \cite{SCM:CPC00} Calculations were carried out for
80 energies, up to 150 K, including all partial waves to convergence. The propagation was
carried out in 2000 log-derivative steps up to a hyperradius of 40 a$_0$. The basis
included all the accessible states up to $E$=3.0 eV.

\subsection{Partial cross sections}
The expression of the  probability for the $j \to j'$ transition and for given
total, $J$, and orbital, $L$, angular momentum values can be written as:
\begin{equation}\label{eq3}
P_{j \to j'}(J,L)=\sum_{L'} \sum_{\epsilon}
|S_{v',j'L'\, \, v,j,L}^{J{\rm \epsilon}}|^2,
\end{equation}
where $S_{v',j'L'\,\,v,j,L}^{J{\rm \epsilon}}$ is the scattering matrix element for the
process between the reactant channel $(v,j,L)$ and the product channel $(v',j',L')$, for
$J$ and the parity $\rm \epsilon=(-1)^{(j+L)}=(-1)^{(j'+L')}$. In Eq.~\eqref{eq3}, the
sums run over the possible values of the final orbital angular momentum $L'$, and the
parity $\rm \epsilon$ .

for a given value of the collision energy, the double partial cross section is given by
\begin{equation}\label{eq10}
\sigma^{J,L}_{j \to j'}=\frac{\pi}{k^2}\, \frac{2J+1}{2j+1} P_{j \to j'}(J,L)
\end{equation}
If summed over $J$ one gets the $J$ partial cross section:
\begin{equation}\label{eq11}
\sum_{J=|L-j|}^{L+j} \, \sigma^{J,L}_{j \to j'}=\frac{\pi}{k^2}\,
\sum_{J=|L-j|}^{L+j} \frac{2J+1}{2j+1} P_{j \to j'}(J,L) =\sigma^L_{j \to j'}
\end{equation}
Alternatively, by summing over $L$, the $J$ partial cross section is retrieved:
\begin{equation}\label{eq12}
\sum_{L=|J-j|}^{J+j} \, \sigma^{J,L}_{j \to j'}=\frac{\pi}{k^2}\,
\frac{2J+1}{2j+1} \sum_{L=|J-j|}^{J+j} P_{j \to j'}(J,L) =\sigma^J_{j \to j'}
\end{equation}

\subsection{1D-model:}

To understand the nature of the peaks observed in the cross-sections,
and confirm their resonant nature, we have used a simple, essentially elastic,
one-dimensional model. It is based on the calculation of 1D adiabatic potentials,
to describe the effective interaction felt by the diatom when it gets
close to the H atom. It is convenient to use
$(R,r,\gamma)$ Jacobi coordinates to describe the scattering process: $R$ is the distance between H and the
centre-of-mass of HF, $r$ the HF internuclear distance (kept fixed in the rigid rotor
approximation) and $\gamma$ the angle between $R$ and $r$.
The potentials, which depend on the initial quantum numbers of the colliding partners,
have been calculated by adiabatically separating $\gamma$ from $R$.
For a given value of $R$, the matrix elements of all the terms in the rigid-rotor
Hamiltonian, except for the radial collision kinetic energy, are calculated in a basis
of reactant channels, labeled by the set of quantum numbers $(J,M,j,L)$ and the parity
$\epsilon$.
Only states with the same values of $J,M$ and $\epsilon$ are coupled.
This way, we diagonalize the block matrix corresponding to each pair $(J,\epsilon)$
($M$ being irrelevant) including as many different $j$s and $L$s as needed for
convergence.\cite{PhysRevLett.109.133201}
The resulting potentials,each of them correlating with a particular set of initial quantum numbers,
 support quasibound states, whose energies lie very close to the observed peaks.

To predict the approximate position, $E_0$, and the width, $\Gamma$,  of the resonances, we calculate the
time-delay function,  $t_{\rm delay}(E)$. The time-delay is (in our context) a measurement  of the time
that the internuclear complex is trapped in the potential well. It can be calculated from the
phase-shift of the 1D continuum wavefunction using the expression:
\cite{child:book,Goldberger:book}
\begin{equation}
  t_{\rm delay}(E)=2 \hbar (d\phi/dE),
\end{equation}
The time-delay functions provided by the 1D model are depicted in Fig.\ref{Fig4}.
It is interesting to note their Lorentzian character. Indeed,
in the absence of background, the time-delay has a pure
Lorentzian line-shape centered precisely at
$E_0$, \cite{Smith:PR1960,Smith:PR1963}
\begin{equation} \label{tdel}
t_{\rm delay}(E)=\hbar \Gamma / [(E-E_0)^2 + (\Gamma/2)^2 ]
\end{equation}
where $\Gamma$ is the FWHM of the line-shape. Accordingly, $\Gamma$ and $t_{\rm
delay}(E_0)$ are related as:
\begin{equation}
t_{\rm delay}(E_0)=\frac{4 \hbar}{\Gamma},
\end{equation}
Finally, the lifetime of the resonance (average time delay) is given by:
\cite{Smith:PR1960}
\begin{equation}
\tau=\frac{\hbar}{\Gamma}=\frac{t_{\rm delay}(E_0)}{4}.
\end{equation}

These Eqs. would allow to calculate the features of a resonance
starting from the time-delay functions in the absence of background
scattering.
However, in the presence of background scattering (as it is the case),
we need to modify somewhat the fitting function. Assuming a
background phase-shift which changes linearly with energy
around $E_0$ ($\delta_{back}(E)=b+a(E-E_0)$), the maximum of the total
time-delay will still provide $E_0$.
In turn, once $E_0$ is known, the width can
be determined fitting the time-delay to the analytical expression:
\begin{equation} \label{tdel}
t_{\rm delay}(E)=a+\hbar \Gamma / [(E-E_0)^2 + (\Gamma/2)^2 ]
\end{equation}
The resulting positions and widths predicted by the 1D model are shown in Table I of the main text.
They have been compared with those extracted from the full scattering
calculation. Assuming a contribution up to first order in $(E-E_0)$
from the background scattering
to the S-matrix elements,\cite{JBSKAGM:S20}
we have used the following function to fit
the $L$-$J$ inelastic probabilities:
\begin{equation} \label{tdel}
P(E)=P_3[(E-E_0)]/ [(E-E_0)^2 + (\Gamma/2)^2 ]
\end{equation}
where $P_3[(E-E_0)]$ is a third degree polynomial in the variable $(E-E_0)$, whose
unknown coefficients are also given by the fitting process.

Let us finally note that, as can be easily conclude from the data in Table I,
the lifetimes provided by the 1D-model are an upper bound to the full-scattering
ones. Indeed, the adiabatic Hamiltonian suppresses the kinetic couplings
between different adiabatic curves. A resonance supported by one of this curves,
can only decay by tunneling through the centrifugal barrier.
However, under the exact Hamiltonian, resonances are coupled to other adiabatic
states; hence there is an alternative mechanism of decay, which is expected
to disminish the lifetime.

\subsection{Alignment-dependent cross sections:}

Let us define a scattering frame with the $z$ axis along the initial relative velocity
  and the $xz$ plane as the one determined by the initial and final relative velocities.
  Let us note with $\beta$ the polar angle that specifies the direction of the polarization
  vector in the scattering frame.
Following Ref. \citenum{AMHKSA:JPCA05}, the cross sections  for a given $\beta$, is given by:
\begin{equation}
\sigma^{\beta}=\sigma_{\rm iso} \sum_{k}^{2 j} (2 k + 1 ) s^{(k)}_0 P_{k}(\cos \beta) A^{(k)}_0
\end{equation}
where $\sigma_{\rm iso}$ is the unpolarized cross section, $P_{k}(\cos \beta)$ are the Legendre polynomials,
  and $s^{(k)}_0$ and
$A^{(k)}_0$ are the intrinsic and extrinsic polarization parameters. The latter is a
geometrical factor  that, for optical pumping, is given by the  $\langle j 0 , k 0 | j 0
\rangle$ Clebsch-Gordan coefficient. The $s^{(k)}_0$ polarization parameters can be
calculated from the S-matrix as: \cite{AHJAJZ:JPCL12}
\begin{equation}
s^{(k)}_0 = \frac{\pi}{\sigma_{\rm iso}\, k^2}\,\,\sum_J \sum_{\Omega', \Omega}
(2 J + 1)  |S^{J}_{j',\Omega', j, \Omega}|^2
\langle j \Omega , k 0 | j \Omega \rangle
\end{equation}
where $\Omega$ and $\Omega'$ are the  helicities, {\em i.e.}, the projections of $J$ on
the directions of the reactant's approach and the product's recoil, respectively. $\beta$=0$^{\circ}$ (which is equivalent to $\Omega=0$) collisions are preferentially head-on, while $\beta$=90$^{\circ}$ implies side-on encounters.

The values of $\sigma^{\beta}$ ($\beta=0^{\circ},\beta=90^{\circ}$)  relative to $\sigma_{\rm iso}$ are shown in Fig.~S3 for the three
transitions considered in this work,  and in a wide range of collision energies. At the
lowest energies, close to the Wigner limit, $\sigma^{\beta}$ is independent on
the preparation. \cite{AAMSA:JCP2005,JCGBBA:PRL19} In the vicinity of the resonance, head-on collisions
prevail while at energies above 200\,K side-on arrangements lead to higher cross
sections. The preponderance of the $\beta=0^{\circ}$ preparation in the cold regime for
the $j=1,2 \rightarrow j'=0$ transitions (upper panels of Fig.~S3) stems from the fact
that for $j'=$0 there is only one possible parity, $(-1)^J$, which includes $\Omega=0$, and hence the weight
of this projection is higher than that when the two parities contribute to scattering
(the $(-1)^{J+1}$ parity does not include $\Omega$=0), as is the case for  the $j=2
\rightarrow j'=1$ transition. At energies above the 100-300\,K
side-on collisions are preferred, as expected from classical arguments.


%
\begin{figure*}
  \centering
  \includegraphics[width=0.90\linewidth]{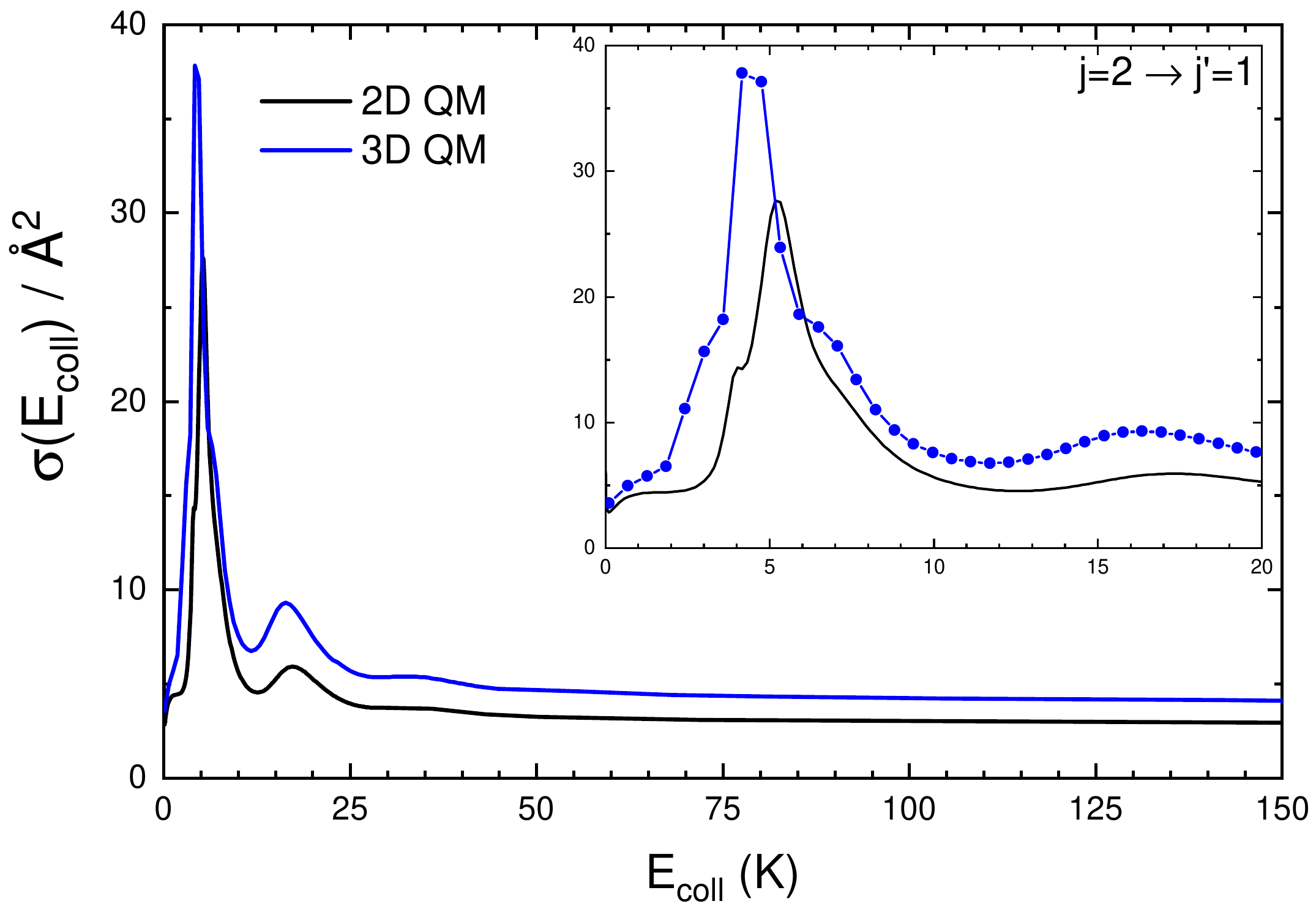}
\caption{ Comparison of 3D and 2D (fixed HF internuclear distance) quantum
scattering calculation for $j=2\rightarrow j'=1$. 2D calculations underestimate the cross
section by 30\% at the resonance peak, but capture all the features observed in the 3D
cross section.} \label{FigS1}
\end{figure*}

\begin{figure*}
\centering
  \includegraphics[width=1.0\linewidth]{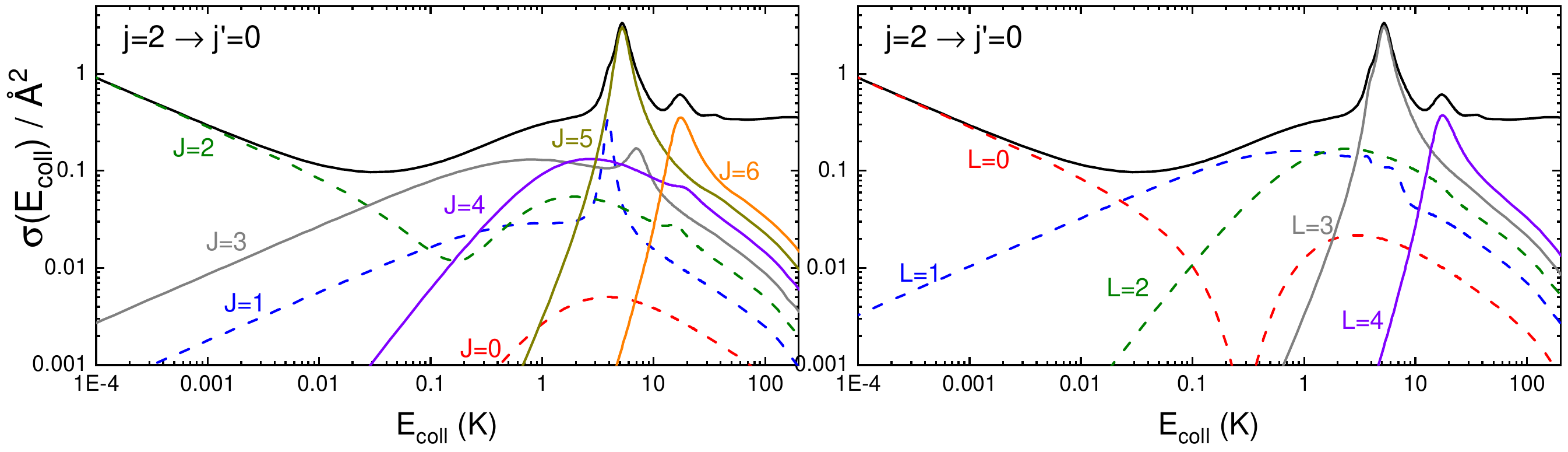}
\caption{ Total (black line), $L$-partial (right panel) and $J$-partial (left
panel) integral cross sections for the $j=2\rightarrow j'=0$ collision.} \label{FigS2}
\end{figure*}

\clearpage

\begin{figure*}
  \centering
  \includegraphics[width=0.8\linewidth]{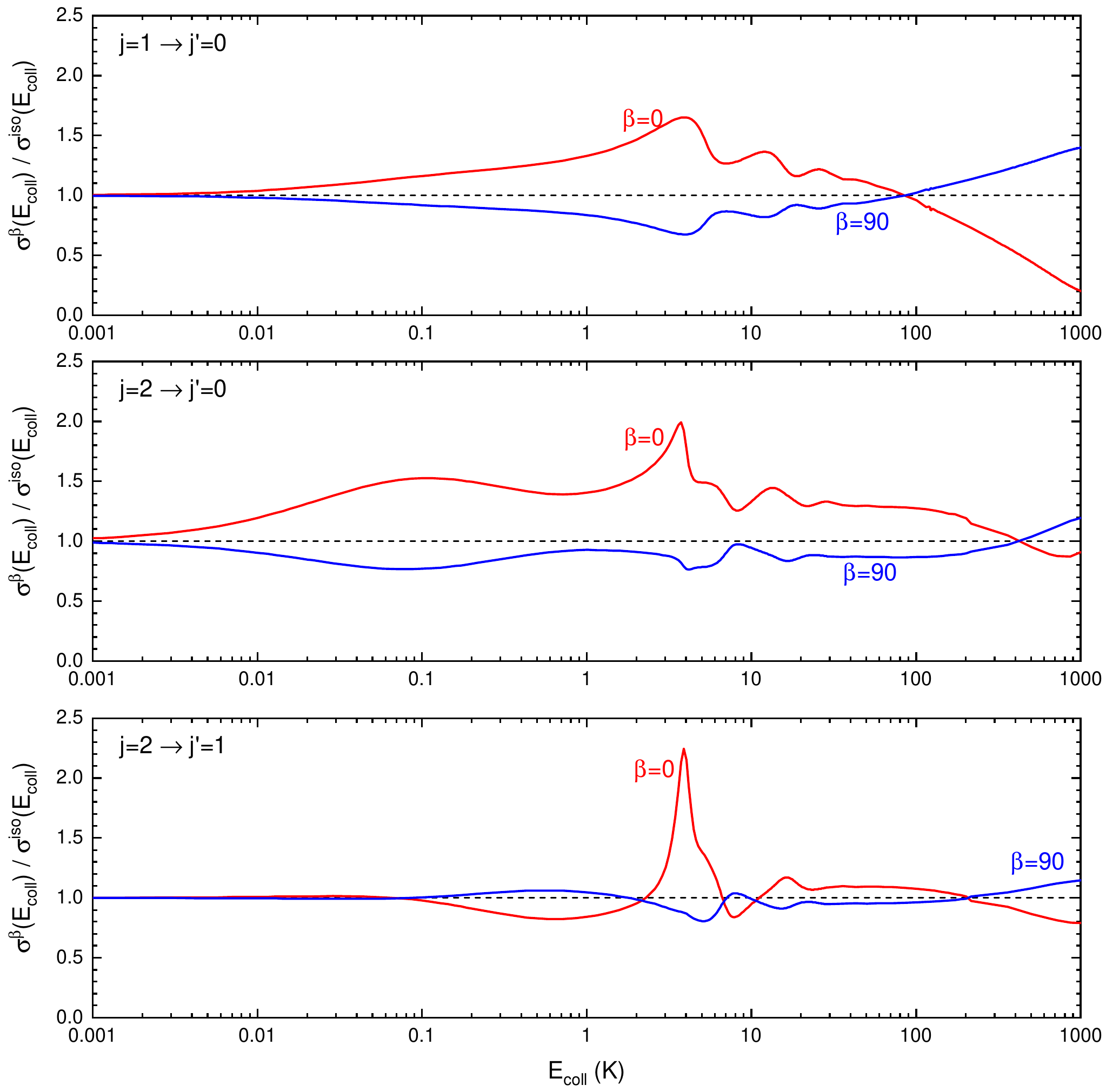}
\caption{ Ratio between the $\sigma(E)$ obtained for $\beta$=0, and
$\beta$=90$^{\circ}$ and the isotropic  $\sigma(E)$. Results for the three opened states
are shown.}
\label{FigS3}
\end{figure*}

\end{document}